\documentclass[aip,
 amsmath,amssymb,
 reprint,%
]{revtex4-1}

\usepackage{graphicx}
\usepackage{dcolumn}
\usepackage{bm}

\usepackage[utf8]{inputenc}
\usepackage[T1]{fontenc}
\usepackage{mathptmx}
\usepackage{multirow}
\usepackage{booktabs}
\usepackage{upgreek}
\usepackage{float}
\usepackage{xcolor}
 
\begin{document}

\title{Band-edge emission enhancement in sputtered ZnO thin films with ultraviolet surface lattice resonances}

\author{Thomas Simon}
\author{Sergei Kostcheev}
\author{Anna Rumyantseva}
\author{Jérémie Béal}
\author{Davy Gérard}
\author{Jérôme Martin}
\email{jerome.martin@utt.fr}
\affiliation{Light, nanomaterials and nanotechnologies (L2n), CNRS ERL 7004, Université de Technologie de Troyes, 12 rue Marie Curie, 10004 Troyes cedex, France}

\begin{abstract}

Metallic nanostructures acting as optical nanoantennas can significantly enhance the photoluminescence (PL) of nearby emitters. Albeit luminescence enhancement factors of several orders of magnitude have been reported for quantum dots or molecules, in the case of bulk emitters the magnitude of the plasmonic enhancement is strongly hindered by the weak spatial overlap between the active medium and the electromagnetic modes of the nanoantenna. Here, we propose a solid-state ultraviolet emitter based on a thin film of zinc oxide (ZnO) coupled with an array of aluminum (Al) nanoparticles. The Al nanorod array is designed to sustain surface lattice resonances (SLRs) in the near ultraviolet, which are hybrid modes exhibiting a Fano-like lineshape with narrowed linewidth relatively to the non-hybridized plasmonic modes. By changing both the period of the array and the dimensions of the nanorods, the generated SLR is tuned either to the near band-edge (NBE) emission of ZnO or to the excitation wavelength. We experimentally demonstrate that NBE emission can be increased up to a factor of 3 compared to bare ZnO. The underlying PL enhancement mechanisms are experimentally investigated and compared with numerical simulations. We also demonstrate that SLRs are more efficient for the ZnO luminescence enhancement compared to localized surface plasmon resonances.

\end{abstract}

\maketitle

\section{INTRODUCTION}

Ultra-compact ultraviolet (UV) light sources are building blocks for key developments in information technologies and biomedical sector. This includes, for instance, laser therapy \cite{surgery}, photocatalysis \cite{photocatalys}, low-threshold solid-state laser sources \cite{Zhang2014}, or high density optical storage \cite{datastorage}. Due to its direct wide bandgap (3.37 eV) and large exciton binding energy (60 meV)\cite{JAPZnO1998}, zinc oxide (ZnO) is a very promising candidate for use in the aforementioned sectors. However, issues linked to the intrinsic properties of ZnO hamper significantly the development of the targeted devices, particularly, if they are meant to operate at the nanoscale. First, the surface recombination at the interface of ZnO results in a shortened carrier lifetime, as the surface recombination through surface/interface states is a very lossy mechanism for photogenerated carriers \cite{ZnOtraps}. This effect becomes much stronger as the geometrical dimensions of the device are reduced due to the increase of the surface to volume aspect ratio.  Second, due to their high refractive index, ZnO structures or thin films induce strong light-trapping, an effect that is not necessarily desired. Third, when used as a gain medium, a ZnO cavity cannot be downsized below half of the wavelength in the considered medium. 

A way to circumvent these issues is to use optical nanoantennas to enhance light-matter interaction at the nanoscale and improve light extraction \cite{Novotny:2011ko}. Optical antennas can be created using metallic nanostructures sustaining localized surface plasmon resonances (LSPRs) \cite{Zayats_2003}. Such resonances are able to confine the electromagnetic energy into deep subwavelength volumes. For that reason, coupling a semiconductor material with a nanostructured metal is a very promising strategy to enhance and/or control the optical properties of the semiconductor at the nanoscale \cite{review2014,Gwo_2016} and to knock down the physical barriers mentioned above. For wide bandgap semiconductors, the metal must sustain good plasmonic properties in the ultraviolet. In that spectral range, one of the best plasmonic materials is aluminum \cite{,Langhammer2008,halas2014,Gerard_2014,Martin_2014}. In contrast with noble metals, aluminum exhibits plasmonic properties in the ultraviolet region, down to a wavelength of 80 nm, while keeping relatively low losses. Moreover, Al is cheap, widely available, and compatible with CMOS technology \cite{Olson:2014cw}. Furthermore, contrary to silver, which also exhibits good optical properties in the near-UV, aluminum is more stable over time, as its oxidation is self-limiting after the formation of a native oxide layer \cite{Zhang:19}. 

In this context, the fabrication of a plasmon-assisted laser with a ZnO nanowire implemented on a single-crystalline aluminum nanometric thin film has been reported \cite{chou2016}. The optical field confinement at the ZnO/Al interface dramatically increases the probability of interaction between the surface plasmon and the gain material, resulting in an enhanced Purcell factor and stimulated excitonic emission. In another study, the ratio between near band-edge emission and defects-related emission from ZnO microrods decorated with Al nanoparticles has been dramatically increased thanks to the resonant coupling with surface plasmons \cite{junfeng2014}. Other studies involve ZnO micro- or nanostructures coupled with Al nanostructures \cite{NOREK201618,sisi2020,ACS2015,JAP2011} or thin films \cite{Zhang2014}. Using ZnO thin films instead of nanostructured ZnO is an alternative worth considering. For instance, Jiang \textit{et al.} reported that an Al metamolecule can selectively enhance the spontaneous emission rate related to the bandgap transition or the defect transition of a ZnO thin film. \cite{ZheyuFang2020} However, the use of a large volume of active material dramatically decreases the spatial overlap between the emitters and the localized electromagnetic field associated with the plasmonic resonances, yielding to small luminescence enhancement factors. 

\begin{figure*}[t]
\includegraphics[width=15cm]{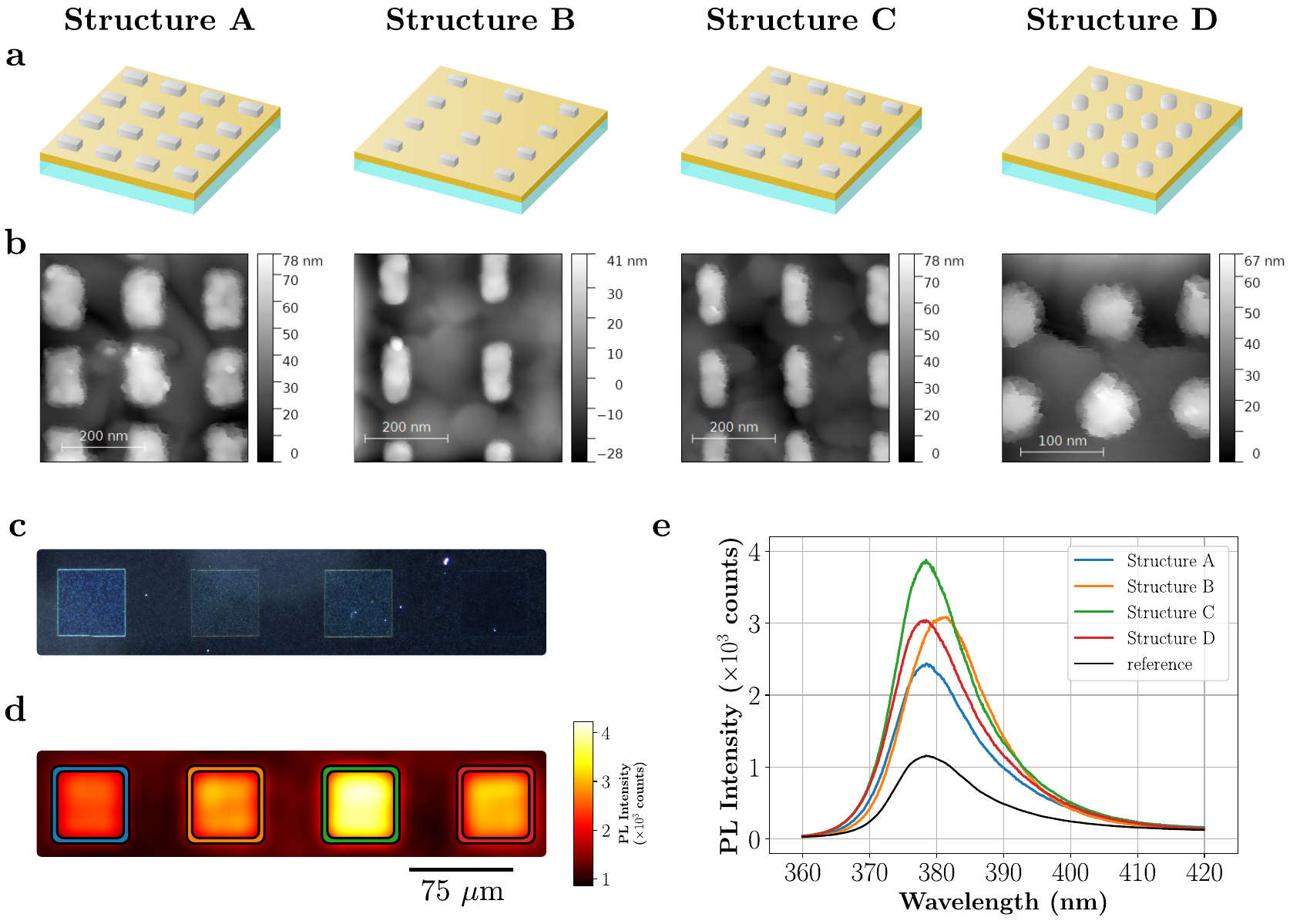} 
\caption{(a) Schematic illustration of the four different Al nanostructure arrays, denoted as A, B, C, and D. The ZnO layer appears in yellow and the quartz substrate in light blue. The geometrical parameters of the arrays are given in Table \ref{samplesprop}. (b) Corresponding AFM images. (c) Dark-field optical microscopy images corresponding to structures A, B, C, and D (from left to right). (d) Maps of PL intensity  at 378 nm over the same area. (e) Corresponding PL spectra. The color of each spectrum corresponds to the color of the box in (d). The reference spectrum corresponds to the average between several measurements on unpatterned areas.}
\label{schemas}
\end{figure*}

A solution to this problem is to hybridize the strongly localized plasmonic modes with \emph{delocalized} photonic modes. Such delocalized photonic modes can be obtained when the nanoparticles are organized into a periodic array exhibiting grazing diffraction orders, or Rayleigh anomalies \cite{zou2004silver, felidj2005grating, auguie08, rivas2009}. The resulting hybrid modes are known as surface lattice resonances (SLRs) or lattice modes \cite{Khlopin17,babicheva2019collective} and manifest themselves as sharp resonances in the extinction spectrum. The electric field intensity associated with the SLR is much more delocalized in the plane of the array than the electric field associated with the plasmonic resonance, a situation much more favorable for solid-state lighting, fluorescence enhancement \cite{PRLRivas2009,murai2021} or infrared plasmonics \cite{murai-IR}.

In this work, we investigate the photoluminescence (PL) properties of a sputtered ZnO thin film combined with an Al nanorod array. The latter exhibits both a narrow plasmonic lattice mode in the UV near the ZnO bandgap and plasmonic resonance in the visible centered on the defect emission wavelength of ZnO. We study both the near band-edge (NBE) emision and the emission from defects in the hybrid structures using spatially resolved micro-PL. When the optical excitation is polarized along the short axes of the nanorods, NBE emission is increased by a factor up to 3 compared to bare ZnO. This effect is attributed to light absorption enhancement allowed by the lattice mode when it matches the excitation wavelength. If the lattice mode is tuned to the NBE emission wavelength, the PL emission is significantly redshifted and the enhancement is much more localized around the lattice mode wavelength. In this configuration, the lattice mode represents a new decay channel for the excitonic emission. With an excitation polarized along the long axes of nanorods, assuming the lattice mode is not excited, we are able to isolate the effect due to the coupling between the ZnO defect emission with the plasmonic mode in the visible. Moreover, Finite Difference Time Domain (FDTD) calculations combined with absorption measurements are used to ascertain the coupling mechanisms. All the measurements and calculations are also done on the reference sample consisting in Al nanorod arrays sustaining non-hybridized LSPRs.

\begin{figure*}[htp]
\includegraphics[width=15cm]{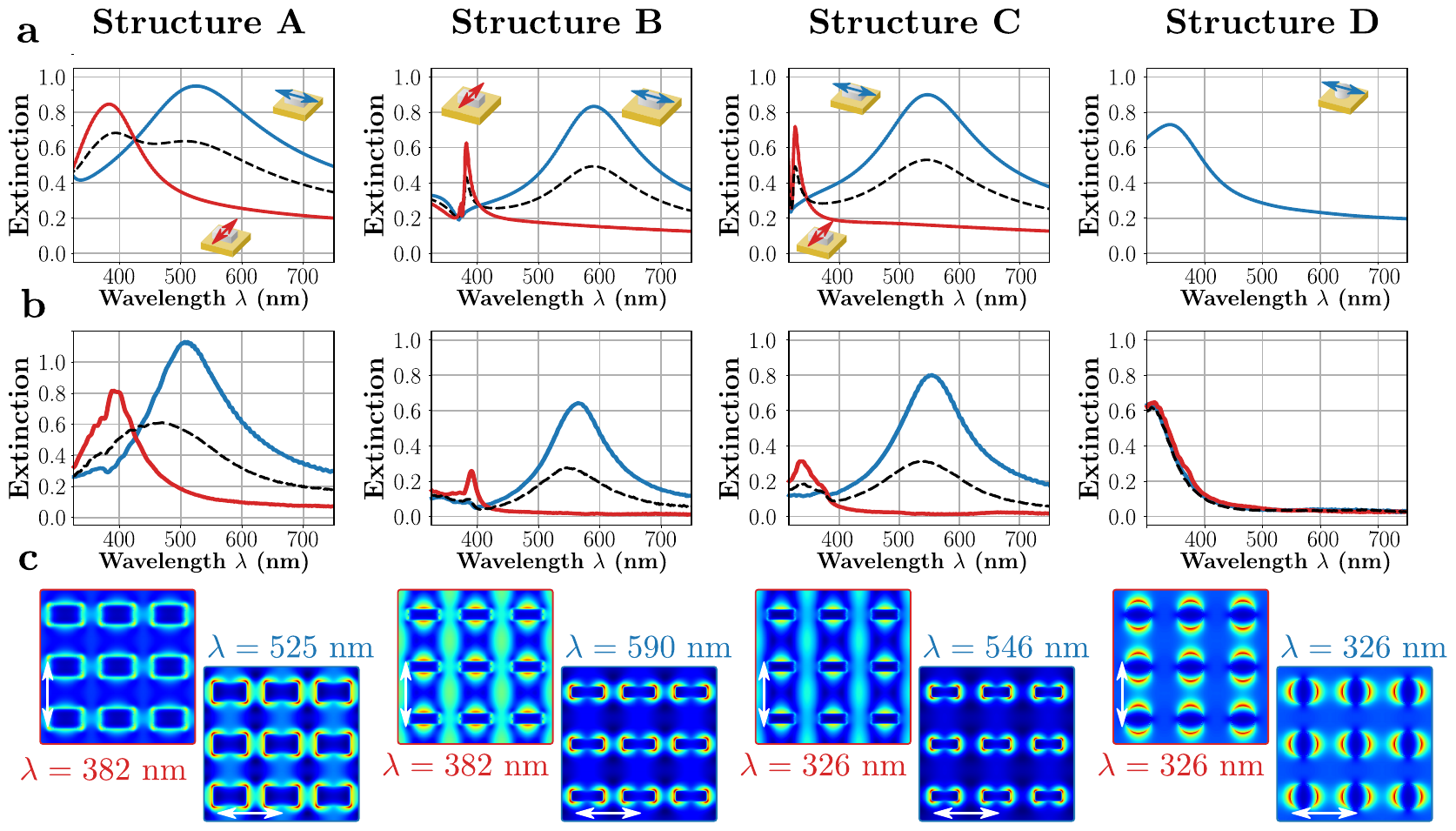}
\caption{Calculated (a) and experimental (b) extinction spectra for both linear polarizations. The dashed lines correspond to the calculated and measured spectra under unpolarized light. (c) Normalized electric field modulus maps at the corresponding resonant wavelengths.}
\label{extinction} 
\end{figure*}

\section{METHODS}

\subsection{Fabrication}

First, ZnO films were deposited onto quartz coverslips by RF magnetron sputtering. A ZnO target with 50 mm diameter was used as the material source. The plasma was activated by a $13.56$ MHz RF power of 200 W at a pressure of $1.0 \times 10^{-2}$ Torr, and the flow of argon and oxygen, respectively, set to 20 and 5 SCCM. The growth rate was approximately 1 nm/min, and the targeted thickness of ZnO set to 40 nm. In a second step, the obtained thin films were subjected to rapid annealing at $1000^{\circ}$C in ambient air during 5 min. This step was mandatory to obtain a steady PL signal (see results in next part) from the thin layer of ZnO. Rapid Thermal Annealing (RTA) of sputtered ZnO films drastically improves their crystal quality and consequently their luminescence properties. \cite{APL-ZnO} The third and last step was the fabrication of Al nanostructures on the top of the ZnO layer, using electron-beam lithography followed by the lift-off procedure.

Figure \ref{schemas}a provides schematic views of the four samples. They consist of square arrays of Al nanoparticles lithographed on a thin ZnO layer. Two geometries were investigated: rods (structures A-C) and cylinders (structure D). This choice was made in order to investigate the effect of polarized illumination on the system, as a single Al nanorod sustains both ultraviolet and visible plasmonic resonances on its short and long axes, respectively \cite{Feifeinanoscale}. The surface topography of the samples was characterized using Atomic Force Microscopy (AFM) as depicted in Fig. \ref{schemas}b. Regular Al arrays were obtained in spite of the relatively high RMS roughness of the underlying ZnO (in the range 6.25 -- 8.37 nm). Also, dark-field optical microscopy images are provided in Fig. \ref{schemas}c, showing the four arrays.

\subsection{Numerical simulations}
Electromagnetic simulations were performed using a commercial software (Lumerical FDTD Solutions). A constant mesh size of 3 nm was set to define precisely the Al nanostructures, while a non-uniform mesh was used outside. A 3 nm layer of Al$_2$O$_3$ was added around the structure to take account of the native oxide layer of aluminum. We used periodic boundary conditions along the $x$ and $y$ axes and perfectly absorbing layers (PMLs) along the $z$ axis of the computation box. The refractive indices of Al and Al$_2$O$_3$ were taken directly from the software's library of materials. ZnO was modeled as a non-absorbing material with a real refractive index $n=2$.

\begin{table}[b]
\begin{tabular}{clccc}
\hline
\begin{tabular}[c]{@{}c@{}}Structure A\\ nanorods\end{tabular}         &  & \begin{tabular}[c]{@{}c@{}}Structure B\\ nanorods\end{tabular} & \begin{tabular}[c]{@{}c@{}}Structure C\\ nanorods\end{tabular} & \begin{tabular}[c]{@{}c@{}}Structure D\\ nanocylinders\end{tabular} \\ \hline
                                                                   &  &                                                            &                                                            &                                                                 \\
L = 130 nm                                                         &  & L = 130 nm                                                 & L = 130 nm                                                 & D = 60 nm                                                       \\
W = 80 nm                                                          &  & W = 50 nm                                                  & W = 45 nm                                                  &                                                                 \\
P = 200 nm                                                         &  & P = 250 nm                                                 & P = 210 nm                                                 & P = 120 nm                                                      \\
                                                                   &  &                                                            &                                                            &                                                                 \\
No SLR                                                             &  & SLR 382 nm                                                 & SLR 326 nm                                                 & No SLR                                                          \\
\multicolumn{1}{l}{}                                               &  & \multicolumn{1}{l}{}                                       & \multicolumn{1}{l}{}                                       & \multicolumn{1}{l}{}                                            \\
\begin{tabular}[c]{@{}c@{}}LSPR 382 nm \\ LSPR 525 nm\end{tabular} &  & LSPR 590nm                                                 & LSPR 546 nm                                                & LSPR 326 nm                                                      \\
                                                                   &  &                                                            &                                                            &                                                                 \\ 
                                                                   
\begin{tabular}[c]{@{}c@{}} $\eta = 2.2$ \end{tabular} &  &            \begin{tabular}[c]{@{}c@{}} $\eta = 3.2$ \end{tabular}  &             \begin{tabular}[c]{@{}c@{}} $\eta = 3.3$ \end{tabular}  &        
\begin{tabular}[c]{@{}c@{}} $\eta = 3.1$ \end{tabular} \\ \hline

\end{tabular}
\caption{Optical properties and geometric parameters of the four structures. All structures have a targeted height of 40 nm. L, W, P, and D stand for length, width, pitch, and diameter, respectively ; $\eta$ stands for the maximum PL enhancement factor in the UV range.}
\label{samplesprop}
\end{table}

\subsection{Optical characterization}
Extinction measurements were performed using a homemade extinction setup. The sample is illuminated by linearly polarized light from a Laser-Driven Light Source (Energetiq). The transmitted light is collected by a NA=0.47 objective lens (LMU-40x-NUV from Thorlabs) and then injected into optical fiber, playing the role of confocal hole, in order to set the size of collection area to $40\, \mu m^{2}$. 

The PL measurements were carried out using a confocal micro-PL bench equipped with a linearly polarized He-Cd laser source emitting at 325 nm. The collection area was roughly equal to 1 $\upmu \mathrm{m}^{2}$. Using motorized translation stages coupled with the sample holder, PL mapping with 5 $\upmu$m spatial resolution has been conducted as shown in Fig. \ref{schemas}e, where the luminescence intensity of ZnO at 378 nm is plotted on a 400 x 100 $\mu m^{2}$ area containing four different Al nanostructure arrays. The ZnO PL enhancement from the four lithographed areas is clearly visible on the maps, as well on the corresponding spectra given in Fig. \ref{schemas}(f). Please note that the presented spectra are restricted to the NBE emission wavelength range. However, there is another emission band located in the visible range and associated with defects in the crystal. This band will be discussed in section \ref{defects}. The geometrical parameters of the arrays and their subsequent plasmonic properties are discussed in Sec. \ref{results}.

\section{RESULTS} \label{results}

\begin{figure}[t]
\includegraphics[width=8cm]{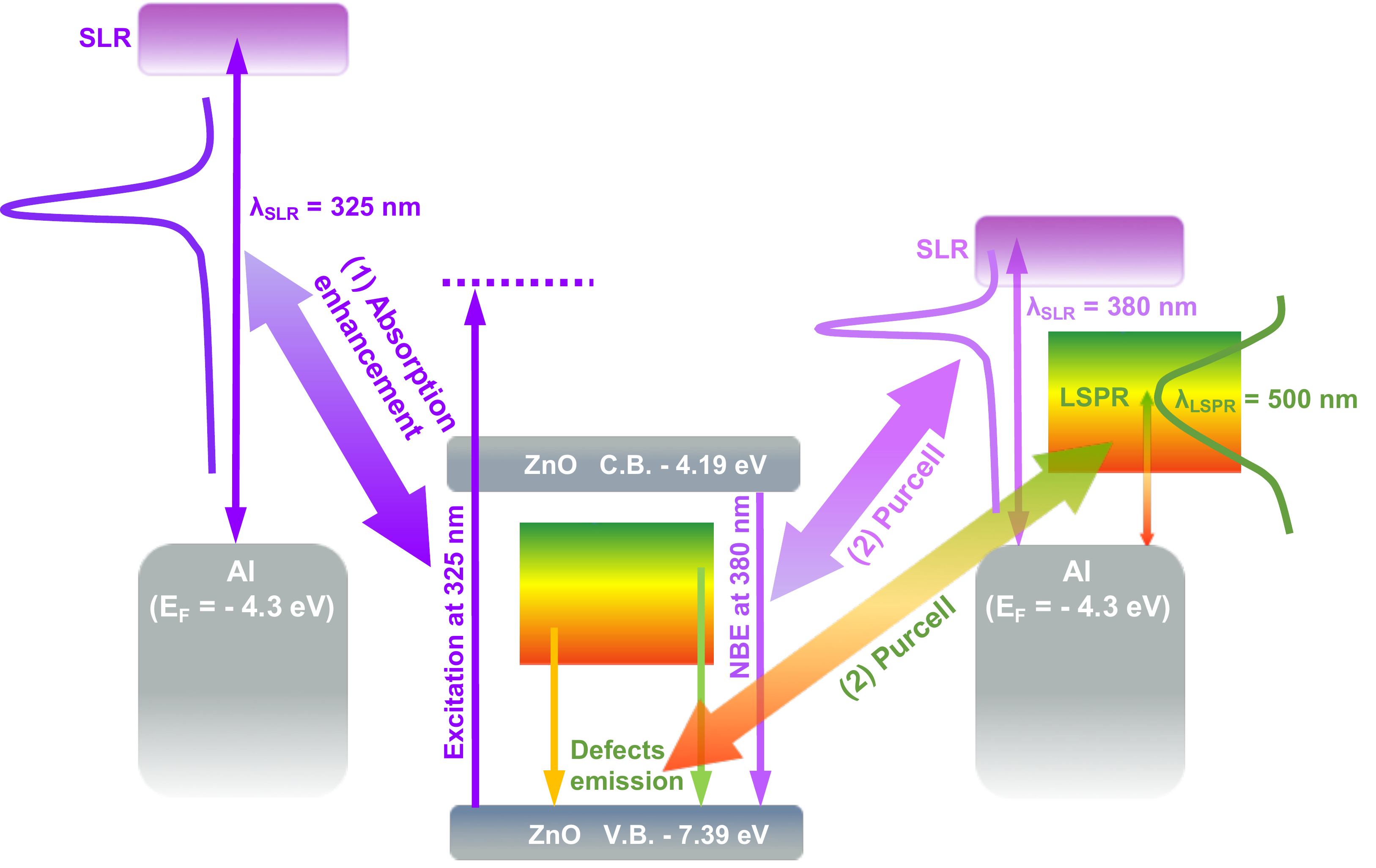} 
\caption{Schematic representation of resonant coupling between SLR and LSPR of Al arrays and the NBE and defects-related optical transitions of ZnO, and the excitation wavelength.}
\label{principe}
\end{figure}

\subsection{Calculated and experimental extinction spectra}

The plasmonic properties of the Al nanostructure arrays have been designed to coincide with the excitation laser source (325 nm) or the NBE emission wavelength (380 nm) and the defect emission (500-600 nm) wavelengths range of ZnO. Four arrays have been fabricated, labeled A, B, C, and D, whose optical properties and geometric parameters are given in Table \ref{samplesprop}. Figure \ref{extinction} shows, for each array, a schematic illustration of the Al array (first row), the calculated (second row), and experimental (third row) extinction spectra, and the normalized electric field modulus maps at the wavelengths of interest (fourth row). All spectra have been calculated and measured under both unpolarized and linearly polarized white light illumination, the polarization being aligned along the short or long axis of the nanorods. The linewidth narrowing experienced by the LSPR in the near ultraviolet when coupled with the $(\pm 1,0)$ Rayleigh anomaly becomes obvious by comparing the spectra of sample A (no SLR) with samples B and C (SLR). The existence of SLRs sustained by samples B and C is ascertained by the electric field modulus maps at 382 and 326 nm respectively: the vertical fringes are associated with a stationary wave corresponding to the interference of the $(\pm 1,0)$ Rayleigh anomalies \cite{Khlopin17}. Note that all nanorod samples also sustain a broad LSPR centered in the visible. Finally, sample D consists of a nanocylinder array exhibiting a single LSPR mode centered at $\lambda=326$ nm. It will be used as a reference sample with no polarization dependence. In the following, we turn our attention to the capability of the hybrid structures to affect the photoluminescence from the ZnO thin layer beneath the Al arrays.

\subsection{Photoluminescence enhancement}

\begin{figure}[t]
\includegraphics[width=\columnwidth]{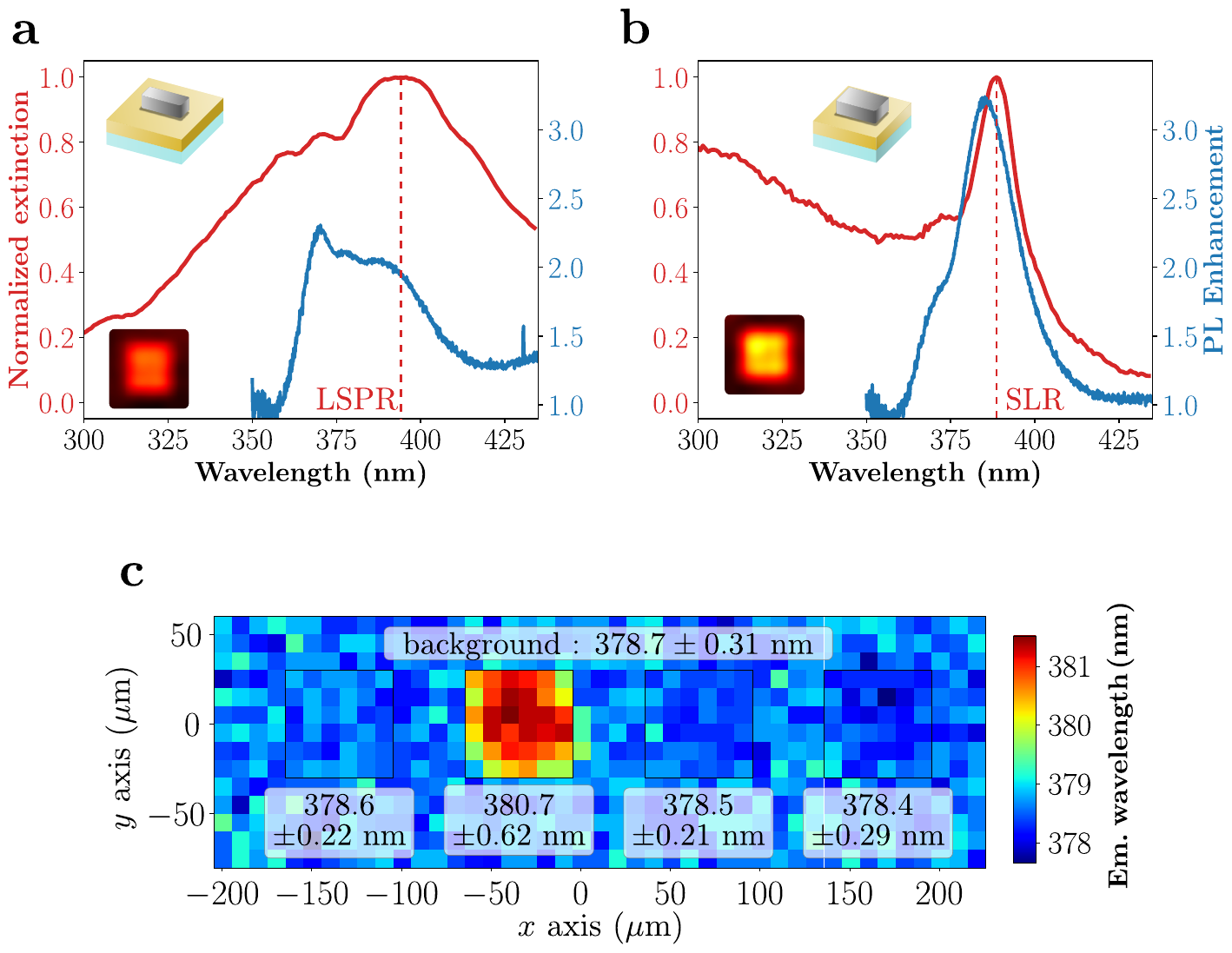}
\caption{(a) Measured extinction spectrum and PL enhancement factors from array A. Inset: PL intensity map measured from the array. Note that the extinction spectrum has been normalized with respect to its maximum value. (b) Same for array B. (c) Map of the spectral PL peak position extracted from the PL measurements.}
\label{carto1um1}
\end{figure}

\begin{figure*}[htp]
\includegraphics[width=11cm]{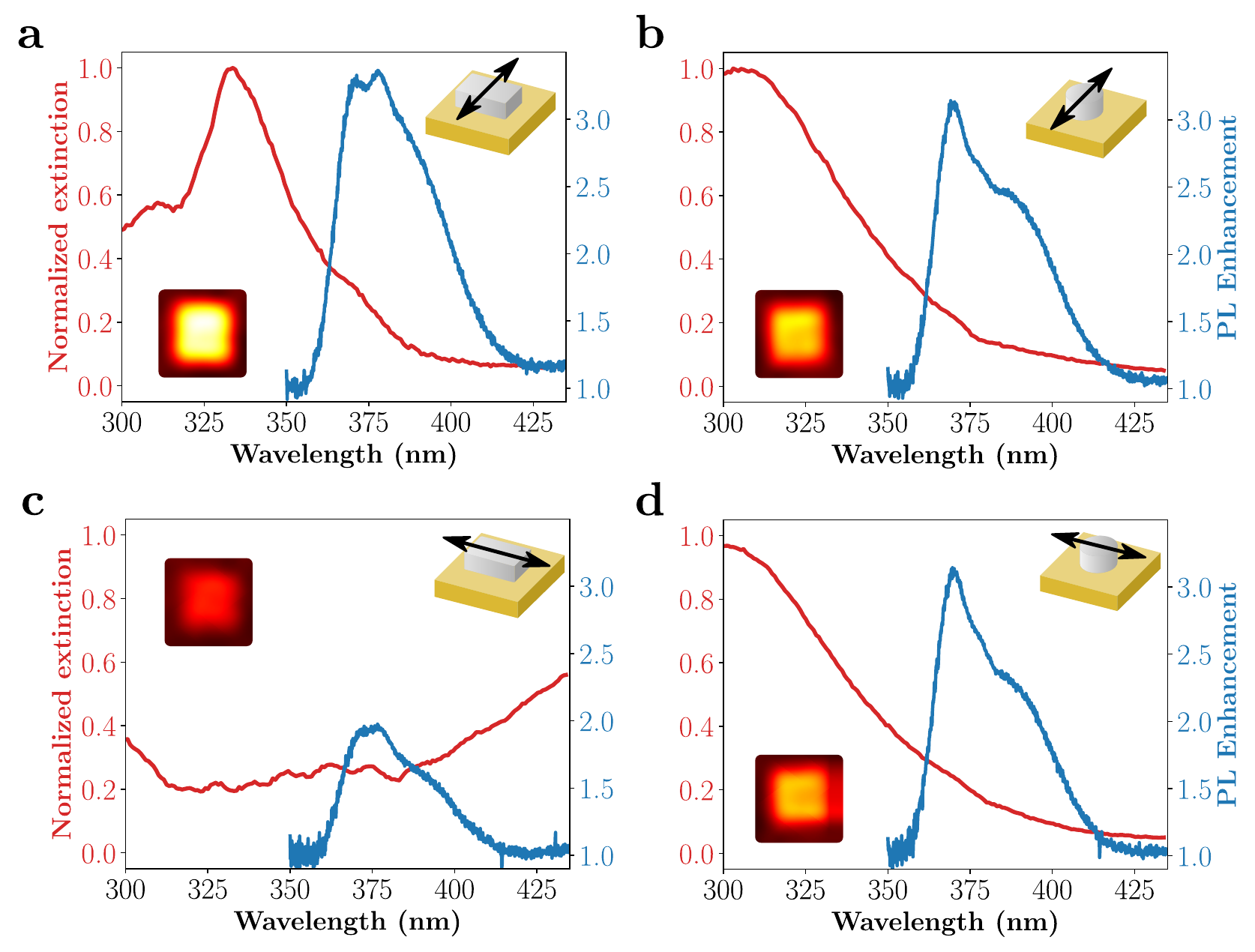}
\caption{Normalized extinction and PL enhancement spectra measured from structures C [(a) and (c)] and D [(b) and (d)] plotted as a function of the wavelength for two polarization states of the impinging laser beam (as indicated by the black arrows in the top right insets). Left insets: PL maps measured from the corresponding array. Note that in (c), the extinction maximum appears for wavelengths longer than the plotted range.}
\label{polar}
\end{figure*}

A metallic nanostructure acting as an optical antenna can alter PL emission in several ways \cite{giannini11b,Goffard13}, as sketched in Fig. \ref{principe}. First, it can increase the local electric field intensity at the pump wavelength (excitation enhancement, see the left part of Fig. \ref{principe}). Then, it can open new decay channels for the nearby emitters, improving the emission rate (the Purcell effect, right panel in Fig. \ref{principe}). If these new channels are radiative channels, then the emission can be enhanced, whereas it will be quenched by non-radiative decay channels. Finally, the antenna can redirect and/or beam the emission, sending more light towards the detector. In the following, we analyze the PL emission from our four structures, using their different properties to disentangle these three effects.

\subsubsection{SLR vs. LSPR for emission rate enhancement}
We first focus on the PL measured on structures A and B. Both of them sustain a plasmonic resonance at $\lambda=382$ nm, very close to the NBE emission of ZnO, except that structure A does not sustain a lattice mode, whereas structure B does. The polarization direction of the impinging light is set along the short axis of the nanorods. Both extinction and PL enhancement obtained from the two arrays are plotted as a function of the wavelength in Fig. \ref{carto1um1}(a,b). The PL enhancement has been calculated as the ratio between the PL measured on the hybrid area containing the Al nanostructures and the PL measured on the bare ZnO next to the structure. The PL enhancement from ZnO coupled with SLR reaches a maximum value of $3.2$  and lies within a narrow range of wavelengths highly correlated with the narrow extinction signature of the SLR. Structure A exhibits a lower PL enhancement, with a peak enhancement of $2.2$ and is distributed over a larger range of wavelengths, linked to the corresponding LSPR extinction signature. We therefore conclude that SLR are more efficient than LSPR in enhancing ZnO photoluminescence in a relatively narrow spectral region centered around the NBE emission of ZnO.

To unveil the underlying mechanisms of the PL enhancement, the peak wavelength of the NBE emission has been mapped as depicted in Fig. \ref{carto1um1}c. The emission peak for structure B is clearly red-shifted compared to all other arrays and to bare ZnO. This effect was actually already visible in Fig. \ref{schemas}e. We emphasize that this spectral shift has been systematically observed in our experiments and does not depend on the numerical aperture of the collecting objective lens. We attribute this effect to luminescence spectral shaping \cite{PRLRivas2009}, where the PL emission in the vicinity of plasmonic arrays is strongly altered. The photonic, delocalized nature of SLR, together with the enhanced electric field between the nanostructures, allow a large coupling efficiency between the luminescent layer and the electromagnetic mode. Consequently, the photogenerated excitons preferentially decays radiatively into SLR modes, exhibiting an enhanced luminescence. As the SLR spectral position ($\lambda=382$ nm) is slightly shifted with respect to the NBE ($\lambda=378$ nm), the hybrid emitter exhibits a spectrally shifted emission at $\lambda=380$ nm, corresponding to a trade-off between both wavelengths.

Finally, we want to emphasize that the two Al arrays discussed here both exhibit a non-zero absorption at 325 nm, the wavelength of the excitation source. Therefore, even if not optimized, the absorption of the excitation light is also enhanced in the vicinity of Al arrays. This will be discussed in Sec. \ref{abs_e}.

\subsubsection{Absorption enhancement and polarization dependence} \label{abs_e}

\begin{figure}[t]
\includegraphics[width=8cm]{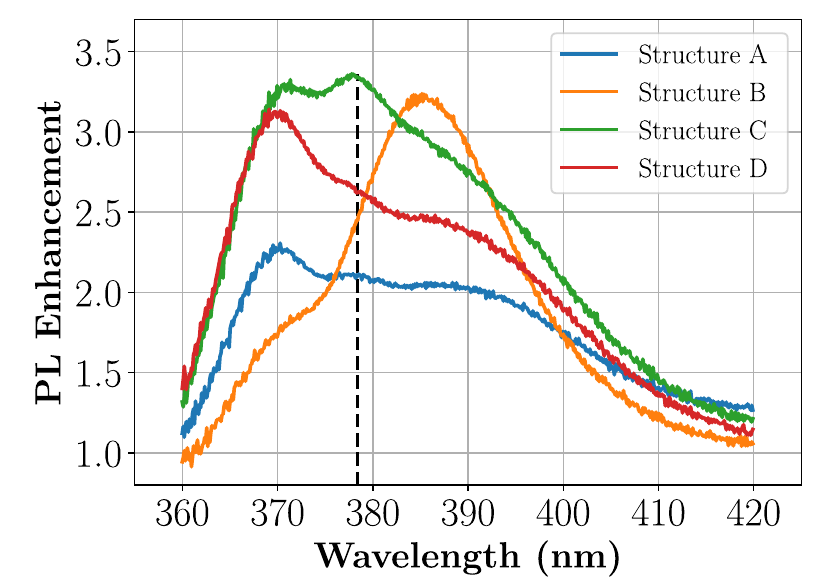}
\caption{PL enhancement factors in the near ultraviolet region plotted as a function of the emission wavelength. The vertical dashed line indicates the ZnO bandgap wavelength. The excitation is linearly polarized and, for structures A-C, is along the short axis of the nanorods.}
\label{enhancementUV}
\end{figure}

We now study PL as a function of the linear polarization state of the impinging laser beam (aligned either along the short or the long axis of the Al nanorods). Moreover, the Al arrays studied in this part are tuned to the excitation wavelength, $\lambda=325$ nm. Results are shown in Fig. \ref{polar}, where PL enhancement and extinction spectra are plotted for structures C and D. It is worth recalling that array C sustains both a SLR at 326 nm (short axis, excitation source matching) and a LSPR located in the visible range (long axis, close to the ZnO defects emission) and that array D sustains only a LSPR at 326 nm (nanocylinder), being used as the reference sample with no polarization dependence. The latter is, therefore, tuned at the excitation wavelength whatever the polarization illumination, which explain why the PL enhancement shown in Fig. \ref{polar}b,d does not depend on the polarization of the impinging laser beam. Polarization dependence is, however, very pronounced on structure C as shown in Fig. \ref{polar}c. 
The enhanced local electromagnetic near-field associated with surface plasmon resonances is known to enhance light absorption by emitters, locally increasing the pump intensity. Because the electron/hole pair generation rate is proportional to this pump intensity, the local concentration of the photogenerated carriers is, therefore, increased. Here, the local pump intensity is higher when the laser source is polarized along the short axis of the nanorods, explaining the observed polarization dependence. 
This demonstrates that the PL intensity from ZnO can be controlled by the polarization state of the illumination source. Our results are summarized in Fig. \ref{enhancementUV}, which shows the PL enhancement factors for the four structures in the 360-420 nm region as a function of the emission wavelength. Figure \ref{enhancementUV} directly evidences the aforementioned mechanisms: structure C, which supports resonances matching both emission and excitation, enhance the PL emission over a larger spectral range than the other structures. The emitted PL, which corresponds to the integral of this spectrum, is hence larger -- although the maximum enhancement factor is similar for structures B-D. Moreover, for structures resonating at the NBE wavelength (structure A and B), a SLR yields larger ZnO PL than LSPRs. The PL shaping observed in Fig. \ref{schemas}e is also obvious in Fig. \ref{enhancementUV} by comparing the PL enhancement spectrum from structure B with the other structures.

\subsubsection{Comparison between NBE and defects-related luminescence}\label{defects}

\begin{figure}[t]
\includegraphics[width=7cm]{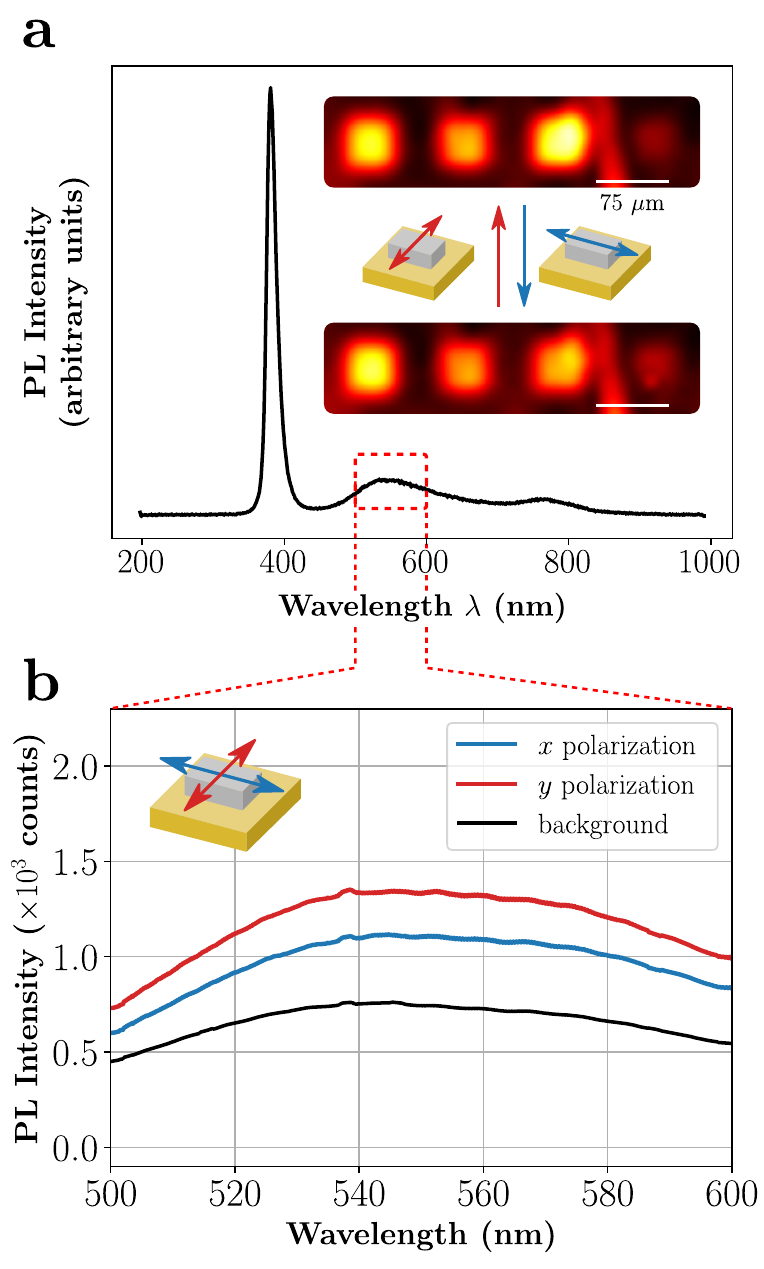}
\caption{(a) Spectrum of annealed ZnO plotted over the UV and visible regions. The small feature around $\lambda = 800$ nm is an experimental artifact (luminescence from the objective). Inset: Maps of PL intensity in the visible range (500 - 600 nm) for the two polarization states of the excitation. The imaged arrays correspond to structures A, B, C, and D (from left to right) from Table \ref{samplesprop}. (b) Zoom in of the defect emission spectral range.}
\label{lumvisible}
\end{figure}

\begin{figure}[t]
\includegraphics[width=7cm]{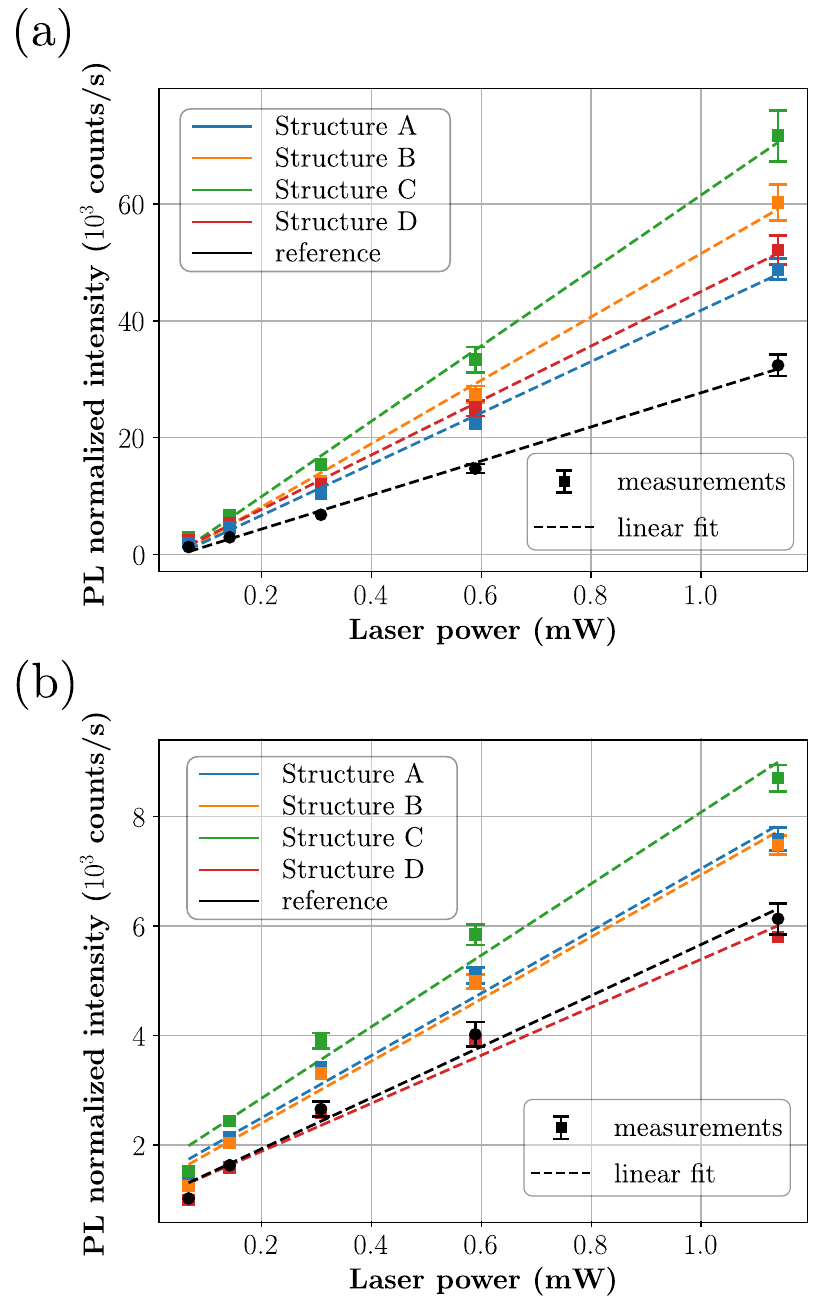}
\caption{Photoluminescence intensity as a function of the pump power: (a) at 380 nm and (b) at 550 nm.}
\label{power}
\end{figure}

Finally, we study the effect of the plasmonic arrays on the luminescence of ZnO defects lying in the visible range. Results are shown in Fig. \ref{lumvisible}a, where maps of the visible PL (500-600 nm) are presented for two polarization states of the incident laser. It appears that all the arrays sustaining a LSPR in the visible region (structures A-C) give rise to the luminescence enhancement (mean value, factor 1.5) from the defects band of the ZnO layer. In contrast, nanocylinders (structure D) do not lead to the significant enhancement of the visible PL. We, therefore, attribute these results to the overlap between the broad LSPR sustained by Al arrays in the visible and the defect emission band. 

Focusing on the emission from defects, the polarization of the source does not affect the PL enhancement except for structure C. Hence, we present in Fig. \ref{lumvisible}b the spectra corresponding to structure C (SLR at 325 nm) for both polarizations. We observe that the enhancement of the visible emission is higher when the SLR at 325 nm is excited, leading to the absorption enhancement.

Figure \ref{lumvisible} also evidences that the PL enhancement is less pronounced for defect-related emission compared to NBE. This is confirmed by the study of the influence of the excitation power on the photoluminescence of the hybrid structures, as shown in Fig. \ref{power}. The PL at 380 nm (a) and 550 nm (b) vs pump power is plotted for bare ZnO and for the hybrid emitters. The linear aspect of the obtained curves indicates that we are operating in the weak excitation regime, where the PL intensity is proportional to the excitation rate \cite{Goffard13}. From Fig. \ref{power}a, the slopes corresponding to ZnO coupled with Al arrays are greater than slopes corresponding to bare ZnO. This indicates that the excitation rate is enhanced for hybrid structures. As expected, the highest enhancement factor is reached for array C (green curves), which SLR is tuned to the excitation wavelength. Regarding the PL from the defects band plotted in Fig. \ref{power}(b), no significant improvement has been observed.

\section{CONCLUSIONS}

In summary, we have studied the optical properties of a ZnO thin layer coupled with Al nanorod arrays. The latter sustain SLRs or LSPRs in the near ultraviolet whether tuned to the excitation source or the ZnO NBE emission wavelengths, and LSPRs tuned to the ZnO defect-related band emission wavelength range. An enhancement of the NBE emission of ZnO up to 3 is demonstrated when coupled with Al arrays. The enhancement mechanisms of NBE emission can be ascribed to (1) the resonant coupling between excitons of ZnO and SLRs and (2) absorption enhancement of ZnO when the SLRs are tuned to the excitation wavelength. When tuned to the NBE emission wavelength, SLRs appear to be more effective to enhance the PL of ZnO compared to LSPRs. We attribute this result to the delocalized and photonic nature of SLRs, allowing for a large spatial overlap of the plasmonic electric field and the semiconductor layer. Due to the intrinsic nanorod anisotropy, we also demonstrate that NBE emission enhancement strongly depends on the polarization direction of the laser source when SLRs are tuned at excitation wavelength. Finally, the visible PL from ZnO is also characterized, and we showed that it can also be enhanced due to the resonant coupling of the Al nanorods LSPRs and the defect-related emission wavelength from ZnO.

\begin{acknowledgments}
We are most grateful to Christophe Couteau and Julien Proust for their help with the experimental setups. TS acknowledges support from the Région Grand Est. Samples were realized on the Nanomat platform (www.nanomat.eu). This work has been done within the framework of the Graduate School NANO-PHOT (École Universitaire de Recherche, grant ANR-18-EURE-0013). This work was supported by the Université de Technologie de Troyes under grant RAMANUV.
\end{acknowledgments}

\subsection*{DATA AVAILABILITY STATEMENT}
The data that support the findings of this study are available from the corresponding author upon reasonable request.

\bibliography{biblio}

\end{document}